\documentclass[a4paper,11pt]{article}
\usepackage[colorinlistoftodos]{todonotes}
\usepackage{placeins}
\usepackage{jheppub} 
\usepackage{units}
\usepackage{siunitx}
\usepackage{graphicx} 
\usepackage{appendix}
\usepackage{amssymb}
\usepackage{slashed}
\usepackage[normalem]{ulem}
\usepackage{bbold}
\usepackage{comment}
\usepackage{lineno}
\usepackage{xcolor}

\newcounter{daggerfootnote}
\newcommand*{\daggerfootnote}[1]{%
    \setcounter{daggerfootnote}{\value{footnote}}%
    \renewcommand*{\thefootnote}{\fnsymbol{footnote}}%
    \footnote[2]{#1}%
    \setcounter{footnote}{\value{daggerfootnote}}%
    \renewcommand*{\thefootnote}{\arabic{footnote}}%
    }
    
\newcounter{afootnote}
\newcommand*{\afootnote}[1]{%
    \setcounter{afootnote}{\value{footnote}}%
    \renewcommand*{\thefootnote}{a}%
    \footnote{#1}%
    \setcounter{footnote}{\value{afootnote}}%
    \renewcommand*{\thefootnote}{\arabic{footnote}}%
    }

\title{First results of the CAST-RADES haloscope search for axions at \unit[34.67]{$\boldsymbol{\mu}$eV}}

\author[1]{A.~\'Alvarez~Melc\'on,}
\author[*,2]{S.~Arguedas~Cuendis,}
\author[3]{J.~Baier,}
\author[2]{K.~Barth,}
\author[4]{H.~Br\"auninger,}
\author[2]{S.~Calatroni}
\author[5,6]{G.~Cantatore,}
\author[2,7]{F.~Caspers,}
\author[8]{J.F.~Castel,}
\author[9]{S.A.~Cetin,}
\author[10]{C.~Cogollos,}
\author[8]{T.~Dafni,}
\author[2]{M.~Davenport,}
\author[11]{A.~Dermenev,}
\author[12]{K.~Desch,}
\author[1]{A.~D\'iaz-Morcillo,}
\author[2]{B.~D\"obrich,}
\author[3]{H.~Fischer,}
\author[2]{W.~Funk,}
\author[13]{J.D.~Gallego,}
\author[1]{J.M.~Garc\'ia~Barcel\'o,}
\author[14,15]{A.~Gardikiotis,}
\author[8]{J.G.~Garza,}
\author[16]{B.~Gimeno,}
\author[11]{S.~Gninenko,}
\author[2,17]{J.~Golm}
\author[18]{M.D.~Hasinoff,}
\author[19]{D.H.H.~Hoffmann,}
\author[8]{I.G.~Irastorza}
\author[20]{K.~Jakov\v{c}i\'c,}
\author[12]{J.~Kaminski,}
\author[5,21,22]{M.~Karuza,}
\author[20]{B.~Laki\'c \daggerfootnote{deceased},}
\author[2]{J.M.~Laurent,}
\author[1]{A.J.~Lozano-Guerrero,}
\author[8]{G.~Luz\'on,}
\author[2]{C.~Malbrunot,}
\author[14]{M.~Maroudas,}
\author[10,32]{J.~Miralda-Escud\'e,}
\author[8]{H.~Mirallas}
\author[23]{L.~Miceli,}
\author[1]{P.~Navarro,}
\author[9,24]{A.~Ozbey,}
\author[9,25]{K.~\"Ozbozduman,}
\author[26,27]{C.~Pe\~na Garay,}
\author[28]{M.J.~Pivovaroff,\afootnote{Current address: SLAC National Accelerator Laboratory, Menlo Park, CA 94025, USA.}}
\author[8,29]{J.~Redondo,}
\author[28]{J.~Ruz,}
\author[8]{E.~Ruiz~Ch\'oliz,}
\author[12]{S.~Schmidt,}
\author[3]{M.~Schumann,}
\author[23,30]{Y.K.~Semertzidis,}
\author[31]{S.K.~Solanki,}
\author[2]{L.~Stewart,}
\author[14]{I.~Tsagris,}
\author[2]{T.~Vafeiadis,}
\author[28]{J.K.~Vogel,}
\author[33]{E.~Widmann,}
\author[2]{W.~Wuensch,}
\author[2,14]{and K.~Zioutas}

\affiliation[1]{Department of Information and Communications Technologies, Technical University of Cartagena, 30203 - Murcia, Spain}
\affiliation[2]{European Organization for Nuclear Research (CERN), 1211 Geneva 23, Switzerland}
\affiliation[3]{Physikalisches Institut, Albert-Ludwigs-Universit\"at Freiburg, 79104 Freiburg, Germany}
\affiliation[4]{Max-Planck-Institut f\"ur Extraterrestrische Physik, Garching, Germany}
\affiliation[5]{Istituto Nazionale di Fisica Nucleare (INFN), Sezione di Trieste, Trieste, Italy}
\affiliation[6]{Universita di Trieste, Trieste, Italy}
\affiliation[7]{European Scientific Institute, Archamps, France}
\affiliation[8]{Center for Astroparticle and High Energy Physics (CAPA) \& Departamento de F\'isica Te\'orica, University de Zaragoza, 50009 - Zaragoza, Spain}
\affiliation[9]{Istinye University, Institute of Sciences, 34396, Sariyer, Istanbul, Turkey}
\affiliation[10]{Institut de Ci\`encies del Cosmos, Universitat de Barcelona
(UB-IEEC), Barcelona, Catalonia, Spain.}
\affiliation[11]{Institute for Nuclear Research (INR), Russian Academy of Sciences, Moscow, Russia}
\affiliation[12]{Physikalisches Institut, University of Bonn, 53115 Bonn, Germany}
\affiliation[13]{Yebes Observatory, National Centre for Radioastronomy Technology and Geospace Applications, 19080 - Guadalajara, Spain}
\affiliation[14]{Physics Department, University of Patras, Patras, Greece}
\affiliation[15]{Universit\"at Hamburg, Hamburg, Germany}
\affiliation[16]{Instituto de F\'isica Corpuscular (IFIC), CSIC-University of Valencia, 46071 - Valencia, Spain}
\affiliation[17]{Institute for Optics and Quantum Electronics, Friedrich Schiller University Jena, Jena, Germany}
\affiliation[18]{Department of Physics and Astronomy, University of British Columbia, Vancouver,Canada}
\affiliation[19]{Xi’An Jiaotong University, School of Science, Xi’An, 710049, China}
\affiliation[20]{Rudjer Bo\v{s}kovi\'c Institute, Zagreb, Croatia}
\affiliation[21]{University of Rijeka, Department of Physics, Rijeka, Croatia}
\affiliation[22]{Photonics and Quantum Optics Unit, Center of Excellence for Advanced Materials and Sensing Devices, and Centre for Micro and Nano Sciences and Technologies, University of Rijeka, Rijeka, Croatia}
\affiliation[23]{Center for Axion and Precision Physics Research, Institute for Basic Science (IBS),Daejeon 34141, Republic of Korea.}
\affiliation[24]{Istanbul University-Cerrahpasa, Department of Mechanical Engineering, Avcilar, Istanbul, Turkey}
\affiliation[25]{Boğaziçi University, Physics Department, Bebek, Istanbul, Turkey}
\affiliation[26]{I2SysBio, CSIC-University of Valencia, 46071 - Valencia, Spain}
\affiliation[27]{Laboratorio Subterr\'aneo de Canfranc, 22880 - Estaci\'on de Canfranc, Huesca, Spain}
\affiliation[28]{Lawrence Livermore National Laboratory, Livermore, CA 94550, USA}
\affiliation[29]{Max-Planck-Institut f\"ur Physik (Werner-Heisenberg-Institut), 80805 - M\"unchen, Germany}
\affiliation[30]{Department of Physics, Korea Advanced Institute of Science and Technology (KAIST),Daejeon 34141, Republic of Korea}
\affiliation[31]{Max-Planck-Institut f\"ur Sonnensystemforschung, 37077 G\"ottingen, Germany}
\affiliation[32]{Instituci\'o Catalana de Recerca i Estudis Avan\c cats, Barcelona,
Catalonia, Spain}
\affiliation[33]{Stefan Meyer Institute for Subatomic Physics, Austrian Academy of Sciences, Vienna, 1030, Austria}

\affiliation[*]{Corresponding author}

\emailAdd{sergio.arguedas.cuendis@cern.ch; babette.dobrich@cern.ch, igor.irastorza@cern.ch, chloe.m@cern.ch}

\abstract{
We present results of the Relic Axion Dark-Matter Exploratory Setup (RADES), a detector which is part of the CERN Axion Solar Telescope (CAST), searching for axion dark matter in the \unit[34.67]{$\mu$eV} mass range. A radio frequency cavity consisting of 5 sub-cavities coupled by inductive irises took physics data inside the CAST dipole magnet for the first time using this filter-like haloscope geometry.
An exclusion limit with a 95\% credibility level on the axion-photon coupling constant of g$_{a\gamma}\gtrsim\unit[4\times10^{-13}]{GeV^{-1}}$ over a mass range of $\unit[34.6738]{\mu eV} < m_a < \unit[34.6771]{\mu eV}$ is set.
This constitutes a significant improvement over the current strongest limit set by CAST at this mass and is at the same time one of the most sensitive direct searches for an axion dark matter candidate above the mass of \unit[25]{$\mu$eV}. The results also demonstrate the feasibility of exploring a wider mass range around the value probed by CAST-RADES in this work using similar coherent resonant cavities.}

\begin{document}
\maketitle
\flushbottom



\section{Introduction}
Astrophysical observations and the standard $\Lambda$ Cold Dark Matter model of the big bang cosmology indicate that around $\sim\unit[84]{\%}$ of the matter in the Universe is dark matter \cite{Ade:2015xua,Bertone:2004pz,Clowe:2006eq}. A suitable candidate for dark matter is the quantum chromodynamics (QCD) axion \cite{Weinberg:1977ma,Wilczek:1977pj}. This pseudoscalar particle was first introduced to solve the strong Charge Conjugation-Parity (CP) problem via the Peccei-Quinn mechanism \cite{Peccei:1977hh,Peccei:1977ur}. Later on it was realized that the axion has the right properties and the proper production mechanism to be a cold dark matter candidate \cite{Abbott:1982af,Dine:1982ah}. Depending on when the Peccei-Quinn symmetry is broken, the production of axions in the early universe can happen before \cite{Tanabashi:2018oca} or after inflation. The post-inflationary scenario predicts axion dark matter masses above roughly \unit[25]{$\mu$eV} \cite{Kawasaki:2014sqa,Fleury:2015aca}.
Benchmark QCD axions follow a strict mass-photon coupling relation. However, variant models allow QCD axions with enhanced coupling to photons \cite{Agrawal:2017cmd,DiLuzio:2016sbl}, and generic axion-like particles (not related to the solution of the strong CP problem) may constitute dark matter with larger couplings than a benchmark axion \cite{Arias:2012az}. A recent review of axion cosmology and the search for axions is given in \cite{Irastorza:2018dyq}.

One way of detecting this particle is by the process in which the axion is converted into a photon in the presence of a strong static magnetic field, the inverse Primakoff effect \cite{Pirmakoff:1951pj,Sikivie:1983ip}. The axion signal power is enhanced using a resonant cavity with a resonance frequency corresponding to the axion mass tested {($\nu=m_ac^2/h$)}. This is known as the haloscope detection method and was first introduced by Pierre Sikivie in 1983 \cite{Sikivie:1983ip}. The expected power extracted from the cavity due to the axion-photon coupling is given by the following equation \cite{Sikivie:1983ip,Krauss:1985ub,Kenany:2016tta}: 

\begin{equation}
\label{eq:axion power}
    P_a = g_{a\gamma}^2\rho_{\text{DM}}\frac{\beta}{1+\beta} \frac{1}{m_a}B^2 V Q_L G^2,
\end{equation}
where $g_{a\gamma}$ is the axion coupling to two photons, $\rho_{\text{DM}} = \unit[0.45]{GeV cm^{-3}}$ \cite{Read:2014qva} is the local dark matter density, $m_a$ is the axion mass, $B$ is the magnetic field strength, $V$ is the volume of the cavity, $Q_L$ is its loaded quality factor, $G$ is a geometric factor that basically represents the overlap of the cavity resonant mode with the magnetic field {and $\beta$ is the coupling between the cavity and the receiver chain}. 

The axion signal generated in the cavity would appear as an increment over the background thermal noise. For this increment to be measurable, the fluctuations of the power measurement must be sufficiently small, which requires long integration times. The noise power in the cavity is given by:
\begin{equation}
\label{eq:Noise power}
    P_N = k_b T_{\text{sys}} \Delta\nu, 
\end{equation}
where $k_b$ is the Boltzmann constant, $T_{\text{sys}}$ is the system noise temperature and $\Delta\nu$ is the resolution bandwidth of our data acquisition system (DAQ). For the scenario where $\Delta\nu = \Delta\nu_a$, Dicke's radiometer equation can be used to set a target signal-to-noise ratio (SNR) \cite{Dicke:1946glx}:
\begin{equation}
\label{eq:SNR}
    \text{SNR} = \frac{P_a \sqrt{t\Delta\nu_a}}{P_N},
\end{equation}
where $t$ is the total integration time, $\Delta\nu_a$ is the axion bandwidth\footnote{{The axion bandwith ($\Delta\nu_a$) is determined by its coherence time $\tau_a=\lambda_a/v\sim \unit[150]{\mu s}$ for an axion mass around $\unit[34]{\mu eV}$ and  $v \sim \unit[270]{km/s}$. $\Delta\nu_a=1/\tau_a\sim \unit[6.6]{kHz}$}} and $P_a$ is the axion power.

Many experiments , see e.g.\cite{Du:2018uak,Braine:2019fqb,Zhong:2018rsr,Lee:2020cfj}, have successfully implemented the haloscope technique to set limits to the axion coupling at low mass ranges (mainly below \unit[25]{$\mu$eV}). Most of the haloscope detectors use a cylindrical cavity in a solenoidal magnet. In order for these detectors to resonate at higher frequency and search for higher axion masses, naively the radius of the cylinder has to be reduced. This, in turn, decreases the total volume of the detector which reduces its sensitivity to the axion signal (see equation (\ref{eq:axion power})). A way to overcome this issue is to create substructures in the cavity to make the cavity resonate at a higher frequency without losing in volume. This is the concept behind our work. A similar idea led to the development of multiple cell cylindrical cavities \cite{Jeong:2017hqs,Jeong:2020cwz}.

This work is based on the idea of using dipole magnets to search for axions using haloscopes \cite{Baker:2011na}. The Relic Axion Dark-Matter Exploratory Setup (RADES), a detector which is installed in the CERN Axion Solar Telescope (CAST)\cite{Anastassopoulos:2017ftl}, uses a new type of cavity geometry where a long rectangular cavity is divided into smaller sub-cavities \cite{Melcon:2018dba} to search for axion masses above \unit[30]{$\mu$eV}. These cavities can resonate at higher frequencies (between 8 and \unit[9]{GHz}) and can be coupled together to increase the volume. This technology can thus, in principle, take advantage of the large volume offered by dipole magnets in order to increase the sensitivity to the axion. Other groups are developing complementary techniques \cite{TheMADMAXWorkingGroup:2016hpc,BRASS} or performed searches at higher masses using new cavity geometries \cite{Alesini:2019ajt,Alesini:2020vny,McAllister:2017lkb,Backes:2020ajv}, but at \unit[34.67]{$\mu$eV} the best limit so far was placed by CAST's solar axion searches \cite{Anastassopoulos:2017ftl}.  

In 2018 the first CAST-RADES prototype was installed inside one of the CAST's magnet bores. In this work the results of the 2018 acquisition campaign are presented. Section \ref{Experimental Setup} briefly outlines the chosen cavity design, its properties and details the experimental setup and the characterization of the quantities involved in equation (\ref{eq:axion power}). Section \ref{Measurements and results} gives an overview of the analysis procedure and presents the results of the measurements. In section \ref{Conclusion and prospects} the results are discussed and prospects for the future are summarized.

\section{Experimental Setup}
\label{Experimental Setup}
\begin{figure}
\centering
\includegraphics[width=0.35 \textwidth]{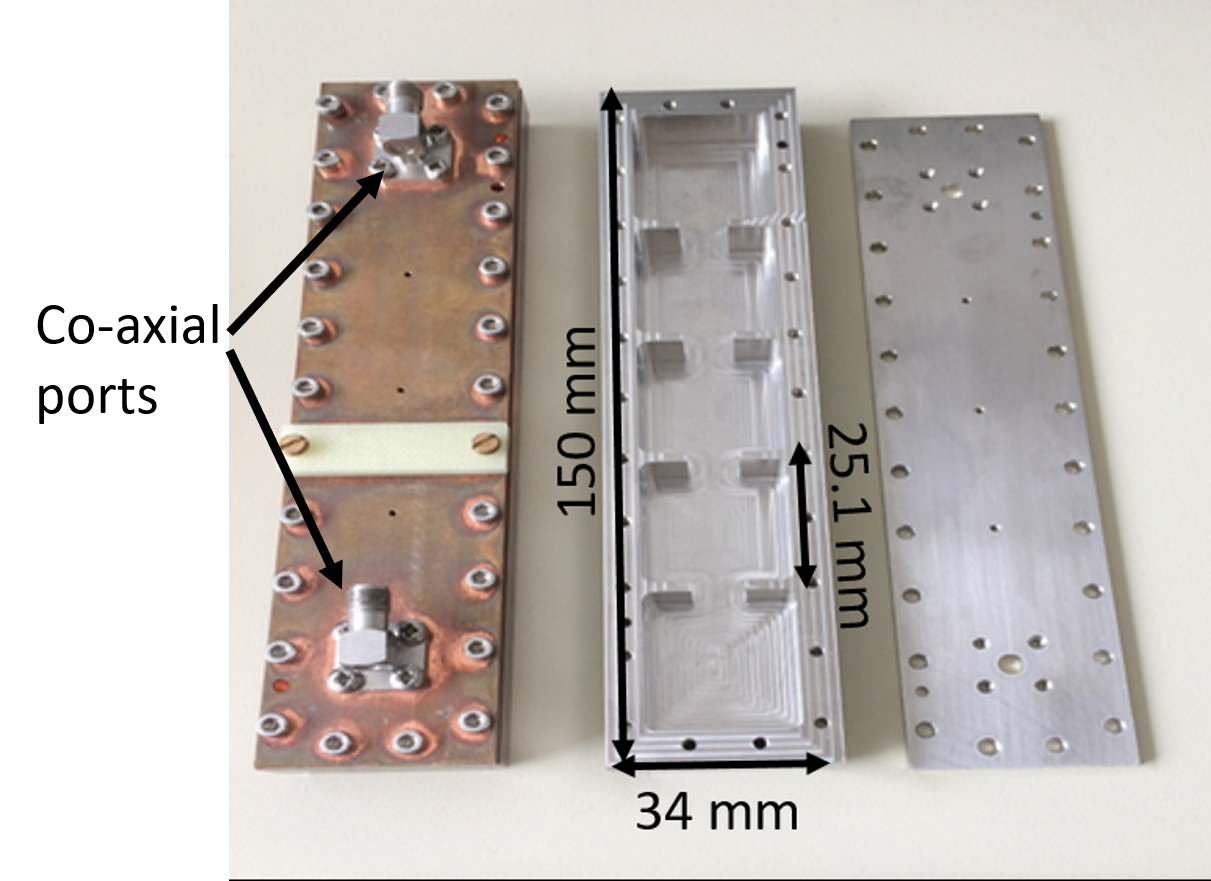}
\includegraphics[width=0.5 \textwidth]{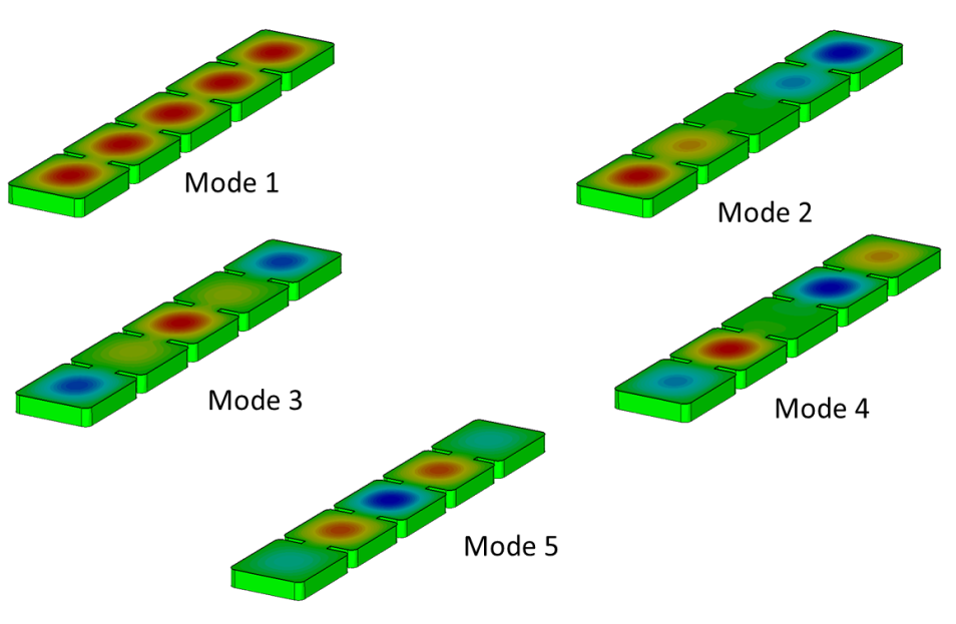}
	\caption{\label{fig:RADES-cavity} 
Left: CAST-RADES  cavity after copper coating (closed) and before the copper coating (opened). Right: Electric field pattern of the cavity modes \cite{Melcon:2018dba}, where red and blue indicate opposite directions of the electric field. The first mode, in which the electric field in all sub-cavities is parallel to the external magnetic field, couples  to the axion.}
\end{figure}

The CAST-RADES detector used in this work consists of a 316LN stainless steel cavity, coated with a $\unit[30]{\mu m}$ thick copper layer. The cavity is internally divided in 5 sub-cavities interconnected by inductive irises, resembling a filter-like structure. A complete description of the cavity can be found in reference \cite{Melcon:2018dba}, here we briefly revisit the main characteristics. The left panel of figure \ref{fig:RADES-cavity} shows a picture of the cavity before and after copper coating.  The right panel of figure \ref{fig:RADES-cavity} shows the electric field pattern of the different resonant modes. The axion couples to the first mode at an axion mass of $\unit[34.67]{\mu eV}$, which corresponds to a frequency of \unit[8.384]{GHz}. This frequency corresponds to the wave-guide dimensions that fit into the CAST cold bore \cite{Melcon:2018dba}. The volume and geometric factor were calculated using the CST simulation software \cite{CST}.

\begin{figure}[h!]
\centering
\includegraphics[width=0.75 \textwidth]{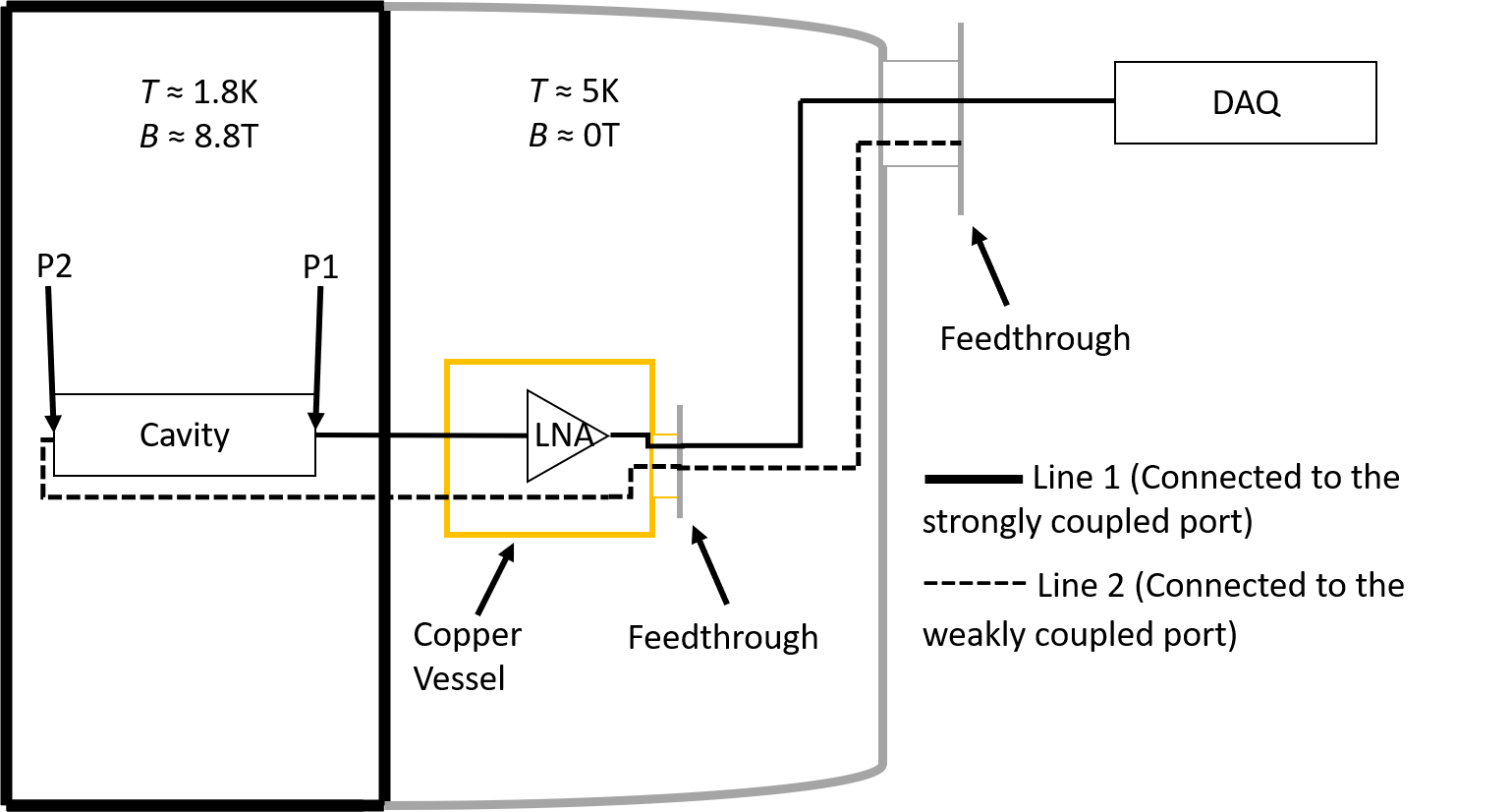}
	\caption{\label{fig:RADES-setup} 
Schematics of the CAST-RADES setup. The cavity is located inside one of the bores of the CAST magnet in a region with $B \approx$ \unit[8.8]{T} at $T \approx$ \unit[1.8]{K}. The critically-coupled port ($P_1$) is connected through a coaxial cable to the LNA, which is located in a region with negligible magnetic field ($B \ll \unit[0.01]{T}$) at $\approx$ \unit[5]{K}. The output of the LNA is connected through coaxial cables to a room temperature feedthrough and then to the DAQ. The weakly-coupled port ($P_2$) is located at the rear part of the cavity and it is connected through coaxial cables to a room temperature feedthrough.}
\end{figure}

The cavity has two 50 $\Omega$ subminiature version A coaxial ports located at each of the extremities, see left panel of figure \ref{fig:RADES-cavity}. One port ($P_1$) is used to extract the signal during the data-taking period, and must be critically coupled to the cavity to maximize the sensitivity. The second port ($P_2$) is used to inject a known input signal to characterize the cavity (e.g. noise temperature or {the frequencies of the cavity resonance modes}). It is very weakly coupled to avoid signal leakage and reduce the noise coming into the cavity from this port. The cavity is placed inside the CAST dipole magnet. Figure \ref{fig:RADES-setup} shows a schematic of the CAST-RADES setup. When the magnet is energized, the magnetic field strength at the position of the cavity is \unit[8.8]{T}. The environmental temperature is \unit[(1.8 $\pm$ 0.1)]{K}. Measurements of the cavity response at cryogenic temperatures but in the absence of a magnetic field have been performed and have served as reference data called magnet-off data ($\mathcal{B}_{\rm off}$).

Port $P_1$ is connected with coaxial cables to a \unit[40]{decibels} (dB) TXA4000 cryogenic low noise amplifier (LNA) manufactured by TTI Norte located in a copper vessel at the end cap of the CAST magnet, in a region with negligible magnetic field ($B \ll \unit[0.01]{T}$), but still at cryogenic temperatures. Both $P_2$ and $P_2$ are connected with semi-rigid coaxial cables to the outside of the cryostat through feedthroughs, as shown in the schematics of Figure \ref{fig:RADES-setup}. The line 1 is then connected to our data acquisition system (DAQ). 

The DAQ system consists of two stages, an analogue and a digital one. The analogue part amplifies further the input signal using a \unit[55]{dB} LNA and down converts it to an intermediate frequency (IF) centered at \unit[140]{MHz} using a local oscillator (LO). The digital stage has a bandwidth of \unit[12]{MHz} with a bin resolution of \unit[4577]{Hz}. The lack of proper isolation between the analog-to-digital converter (ADC) card and the field-programmable gate array (FPGA) resulted in noisy bins in the power spectrum, which were treated at the analysis level (see section \ref{Measurements and results}).

The quality factor $Q_L$, the noise temperature $T_{\text{sys}}$ and the coupling $\beta$ need to be experimentally measured in order to determine the noise power (see equation (\ref{eq:Noise power})) and the corresponding axion-photon coupling to which we are sensitive (see equation (\ref{eq:axion power})). We extracted $Q_L$ from the {DAQ} recorded {power spectra} directly, {in this configuration line 2 is terminated at the room temperature feedthrough with a \unit[50]{$\Omega$} load}. The {power spectra were} recorded with the magnetic field on at two different LO frequencies, which were divided to remove structures that are related to the DAQ electronics\footnote{As visible in the left panel of figure \ref{fig:Spectra}, the power spectrum is not flat outside the main resonance peak and has a non-flat component also convoluted with the main peak. Division of the spectra at different LO frequencies ameliorates this problem, as illustrated on the right panel of figure \ref{fig:Spectra}.}. 
 The divided data was split into sets of 2.5 hours. For each set, the $Q_L$ was obtained by fitting a Lorentzian curve to the measured noise spectrum. The average of the Q-value obtained is $Q_L=11009\pm483$, which was used for the computation of the exclusion limit.

The configuration used during data-taking did not allow for a by-pass of the cryogenic LNA to measure the cavity coupling in-situ. A radio frequency (RF) switch was installed after conclusion of data-taking period upstream of the LNA in order to allow such a measurement without disruption to the magnet's \unit[2]{K} region. The transmission coefficient, $S_{12}$ \cite{Pozar:882338}, was measured with a vector network analyzer (VNA) and yielded $Q_L = 11259$, in agreement with the $Q_L$ measured during the data-taking period. It is therefore justified to assume that the coupling measured after the data-taking period was identical to the cavity coupling during data-taking period.
The reflection coefficient, $S_{11}$ \cite{Pozar:882338}, was also measured using a VNA {, which was connected to line 1. $P_2$ was very weakly-coupled, thus the system can be regarded as a singly terminated resonator,} from which we extracted the coupling $\beta$ using the following relation:
\begin{equation}
    \label{eq:Betap}
    \beta = \frac{1 \mp |S_{11}|}{1 \pm |S_{11}|},
\end{equation}
where $S_{11}$ is in linear units, and where the minus and plus signs in the nominator and denominator, respectively, correspond to an under-coupled configuration (our configuration) and the opposite to the over-coupled configuration. A $\beta$-value = 0.50 $\pm$ 0.11 was calculated and used for the computation of the experimental sensitivity.

For the computation of the noise temperature the Y-method was followed \cite{Y-method}. During the data-taking period, it was not possible to install a noise source in close proximity to the cryogenic LNA. The noise source was thus connected to line 2 of figure~\ref{fig:RADES-setup} at the room temperature feedthrough and the injected noise had to propagate through $\sim$ \unit[5]{m} of cables, two feedthroughs and the cavity before arriving to the cryogenic LNA. Taking into account the added noise of these elements and their uncertainties, we obtained a system noise temperature  of \unit[(7.8  $\pm$ 2.0)]{K} at the resonance peak. 

The coupling $\beta$, noise temperature $T_{\text{sys}}$ and $Q_L$ were inferred from single measurements. They are assumed to have been stable over the data-taking period of 103 hours taken within a period of 20 days. Since our cavity did not undergo any mechanical changes during the data-taking period, the assumption of stability of these parameters is justified.

\section{Measurements and results}
\label{Measurements and results}
\begin{figure}[]
\centering
\includegraphics[width=0.45 \textwidth]{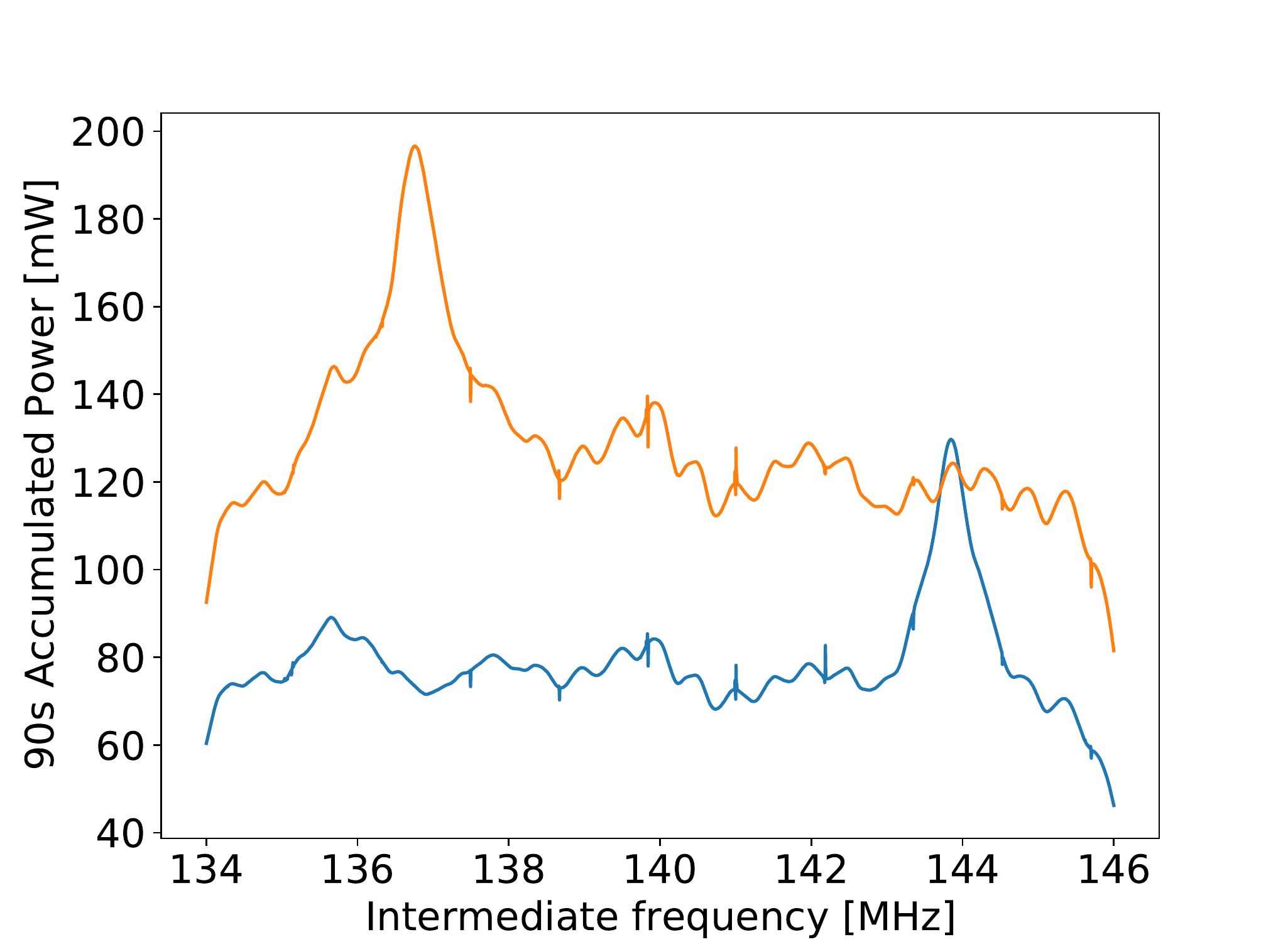}
\includegraphics[width=0.45 \textwidth]{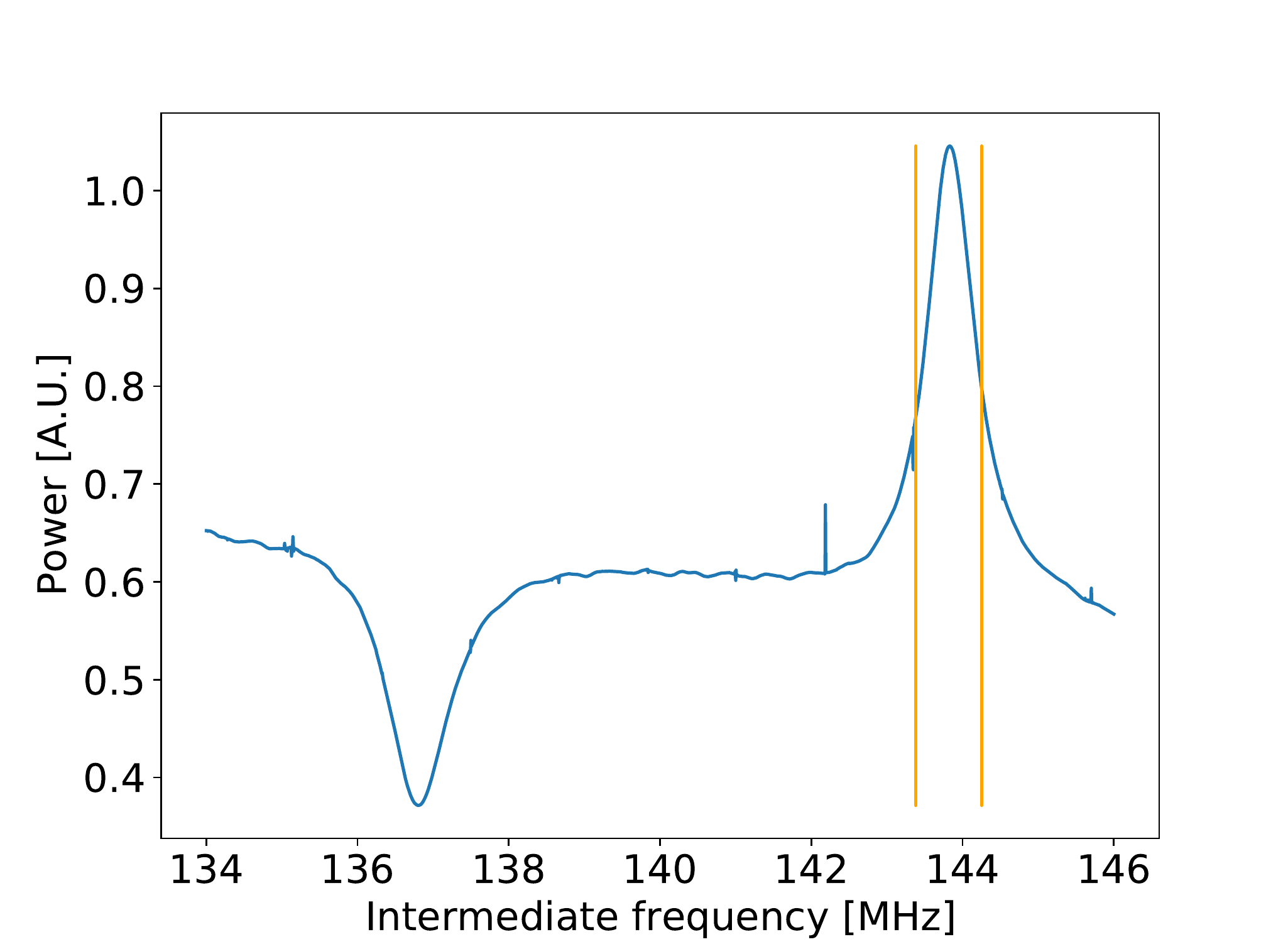}
	\caption{\label{fig:Spectra} 
Left: Typical spectra for two different LO frequencies ($l1 = \unit[8.240]{GHz}$ and $l2 = \unit[8.247]{GHz}$). Upon changing the LO frequency, the cavity resonance peak changes position in the IF frequency. The visible spikes in the spectra are noisy bins produced by the improper isolation between the ADC card and the FPGA. Right: An example of a $\delta_{ij}^d$ spectrum obtained by the division of two spectra taken at two different LO frequencies. This procedure removes the wave-like structures originating from the electronics which are common to both spectra. The orange vertical lines represent the frequency range considered in the analysis which covers a region without noisy bins. {The physical frequency (PF) of the input signal can be calculated using PF = IF + LO.}}
\end{figure}

During the data-taking period a power spectrum was taken every $t = \unit[90.37]{s}$, imposed by the sampling rate of the ADC card and the accumulated fast Fourier transformations made by the FPGA. A total time of about \unit[103]{h} were used for the analysis. Two typical \unit[90]{s} spectra are shown in the left panel of figure \ref{fig:Spectra}. The spectrum consists of the  cavity resonance peak on top of a structure that we have identified to be an electronic background. 

As previously mentioned, in order to isolate the peak of interest from the electronic background, the data was recorded at two different LO frequencies. In the two data sets the peak is displaced in the intermediate frequency (IF), but the electronic background is the same qualitatively, as shown in the left panel of figure \ref{fig:Spectra}. Let us denote the data recorded by $\delta_{ij}$, where $j$ represents the IF bin number  (the total number of bins, 2622, is determined by the total recorded bandwidth divided by the resolution bandwidth) in the $i$-th spectrum, for $i = 1,...,M$ and $j = 1,...,2622$. Here $M = 4093$ is divided in five sets (of 584, 545, 592, 627 and 1745 spectra) of magnet-on data separated by periods of magnet-off data.
.

We will label the spectra taken at the two different LO frequencies  $\delta_{ij}^{l1}$ and $\delta_{ij}^{l2}$. A large part of the electronic background can be removed by dividing each sprectra by the average spectrum of the second LO frequency:

\begin{equation}
\label{eq:Div-Spectra}
    \delta_{ij}^{d} = \frac{\delta_{ij}^{l1}}{\frac{1}{M}\sum_{i =1}^{M} \delta_{ij}^{l2}}.
\end{equation}

We focus the analysis around a limited frequency range located at the Lorentzian peak ($j = 2049,...,2239$) , which corresponds to a frequency range of $\sim$ 0.87 MHz. This range was selected because it covers more than the full width at half maximum of the Lorentzian peak in a region where there are no noisy bins, see {region between the vertical lines in the right panel of figure \ref{fig:Spectra}}. Each IF bin corresponds to a specific physical frequency (PF = IF + LO). The spectra are then aligned so that the bins are labelled according to the physical frequency, for further processing. The index $k$ is reserved to label bins in the PF space.

\begin{figure}[h!]
\centering
\includegraphics[width=0.65 \textwidth]{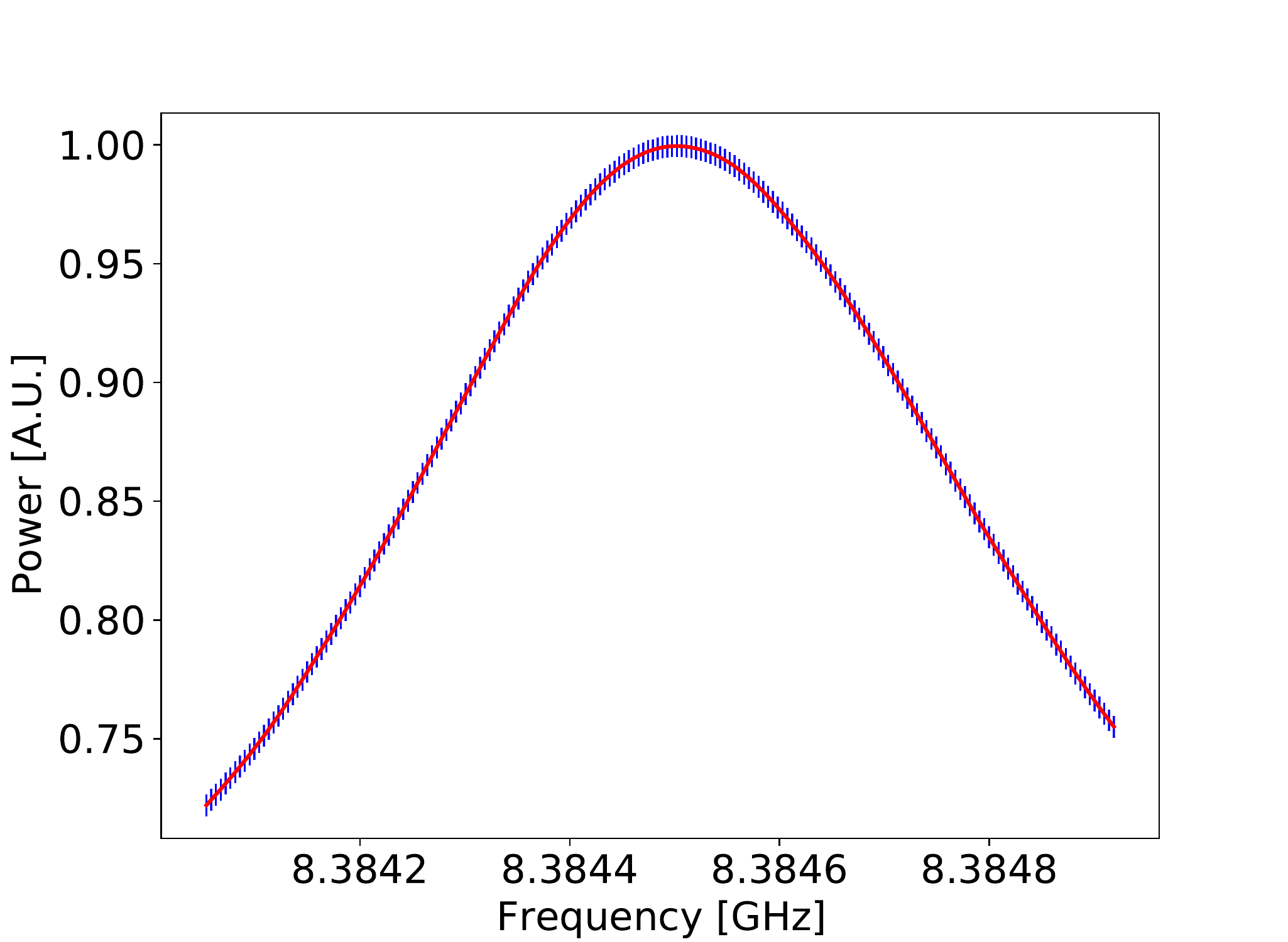}
	\caption{\label{fig:Working Window} 
Typical \unit[90]{s} spectrum  after the division procedure. The blue {vertical lines} indicate the data points and the red line the SG-fit.The plotted frequency range corresponds to the region delimited by the two vertical lines in the right panel of figure \ref{fig:Spectra}.}
\end{figure}

The removal of the remaining structure is largely based on the HAYSTAC analysis procedure \cite{Brubaker:2017rna}. A first Savitzky-Golay (SG) filter \cite{savitzky64} (with 15 points and a polynomial degree of 3) is applied to the average spectrum $\delta^d_k = (1/M)\sum_{i =1}^{M}\delta_{ik}^{d}$, producing a smoothed spectrum called SG-fit. This first SG filter removes the large structure originating from the electronics. {Figure \ref{fig:Working Window} shows a \unit[90]{s} spectrum (blue vertical lines) and the SG-fit (in red) on top of it.} The normalized spectra ($\delta_{ik}^{n}$) are obtained through dividing $\delta_{ik}^d$ by the SG-fit:

\begin{equation}
\label{eq:Norm-Spectra}
    \delta_{ik}^{n} = \frac{\delta_{ik}^{d}}{\text{SG}_k}.
\end{equation}

A second SG-fit ($\text{SG}^*$, produced using a SG-filter with 109 points and a polynomial degree of 3) has to be applied to each $\delta_{ik}^{n}$ to remove the effect of gain drifts and the normalized spectra are once again divided by these fits to obtain Unitless Normalized Spectra:

\begin{equation}
\label{eq:Flat-Spectra}
    \delta_{ik}^{u} = \frac{\delta_{ik}^{n}}{\text{SG}_{ik}^*} - 1.
\end{equation}

\begin{figure}[h!]
\centering
\includegraphics[width=0.5 \textwidth]{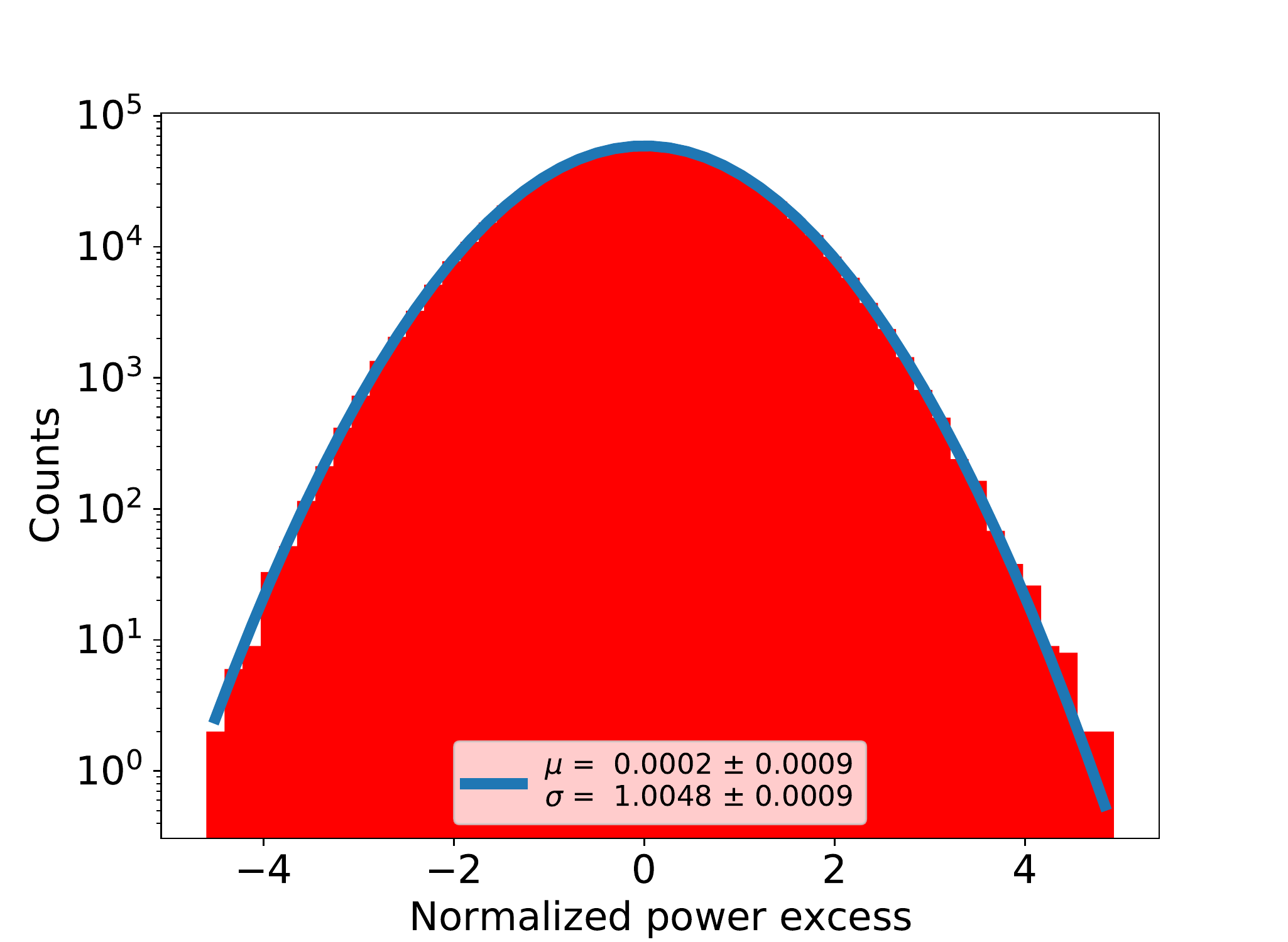}
	\caption{\label{fig:Histos}}
Normalized histogram for all $\delta_{ik}^{u}$.
\end{figure}

In the absence of axion conversion and if appropriate  SG parameters are used, these $\delta_{ik}^{u}$ should be samples of a Gaussian distribution center at $\mu$ = 0 and standard deviation $\sigma$ = $1/\sqrt{\Delta\nu\cdot t}$. Based on the transfer function of the SG filter, we optimized the parameters to achieve the aforementioned Gaussian distribution with the least impact on a possible axion signal. Figure \ref{fig:Histos} shows the histogram for all $\delta_{ik}^{u}$. The $\delta_{ik}^{u}$ are then combined into a Grand Unified Spectrum: 

\begin{equation}
\label{eq:GUS-Spectra}
    \delta_{k}^{g} = \frac{1}{M}\sum_{i =1}^{M} \delta_{ik}^{u},
\end{equation}
with the sample variance:

\begin{equation}
\label{eq:GUS-Spectra-sigma}
    (\sigma_{k}^{g})^2 = \frac{1}{M-1}\sum_{i =1}^{M} (\delta_{ik}^{u} - \delta_{k}^{g})^2.
\end{equation}

\begin{figure}[h!]
\centering
\includegraphics[width=0.45 \textwidth]{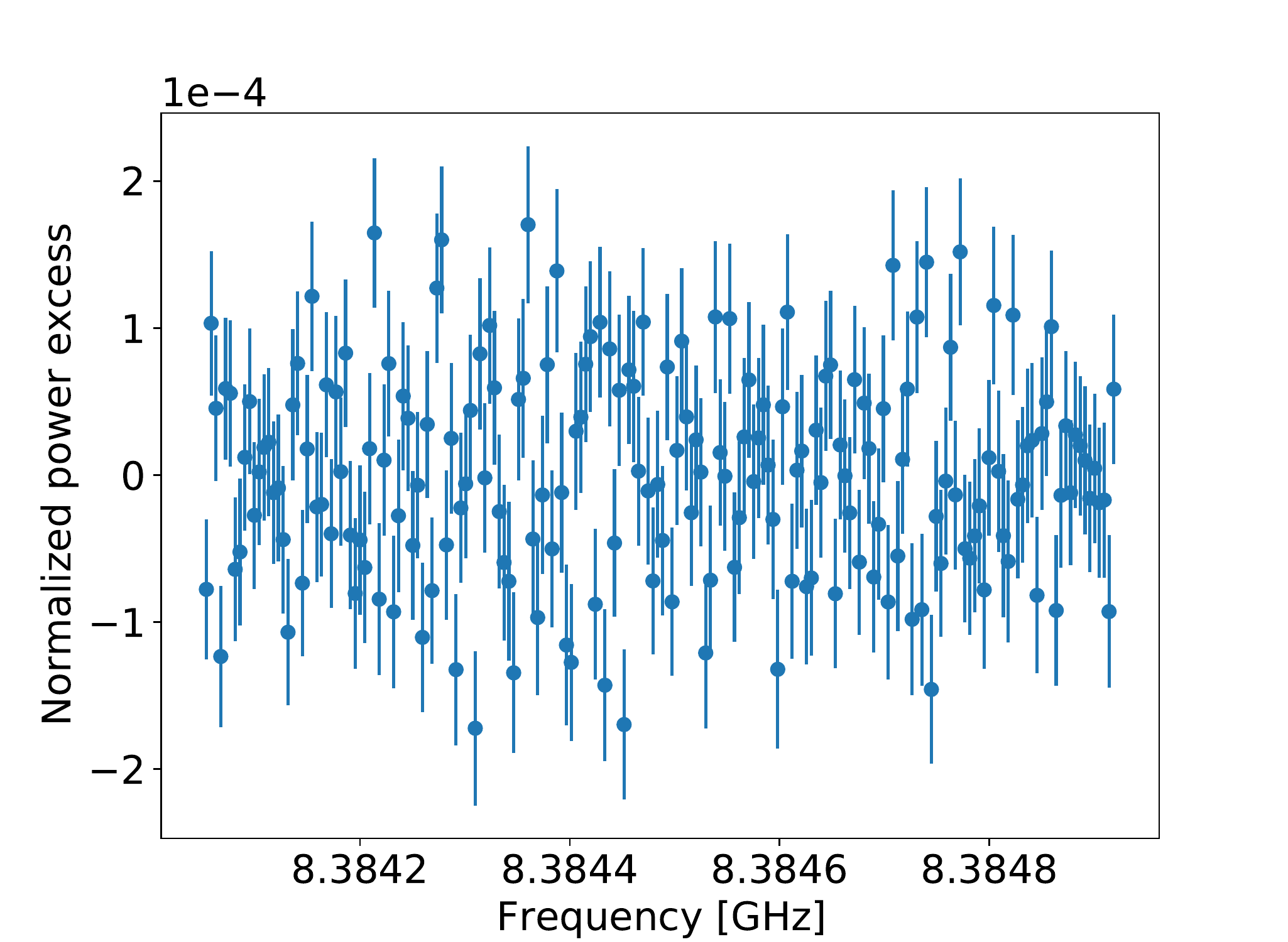}
\includegraphics[width=0.45 \textwidth]{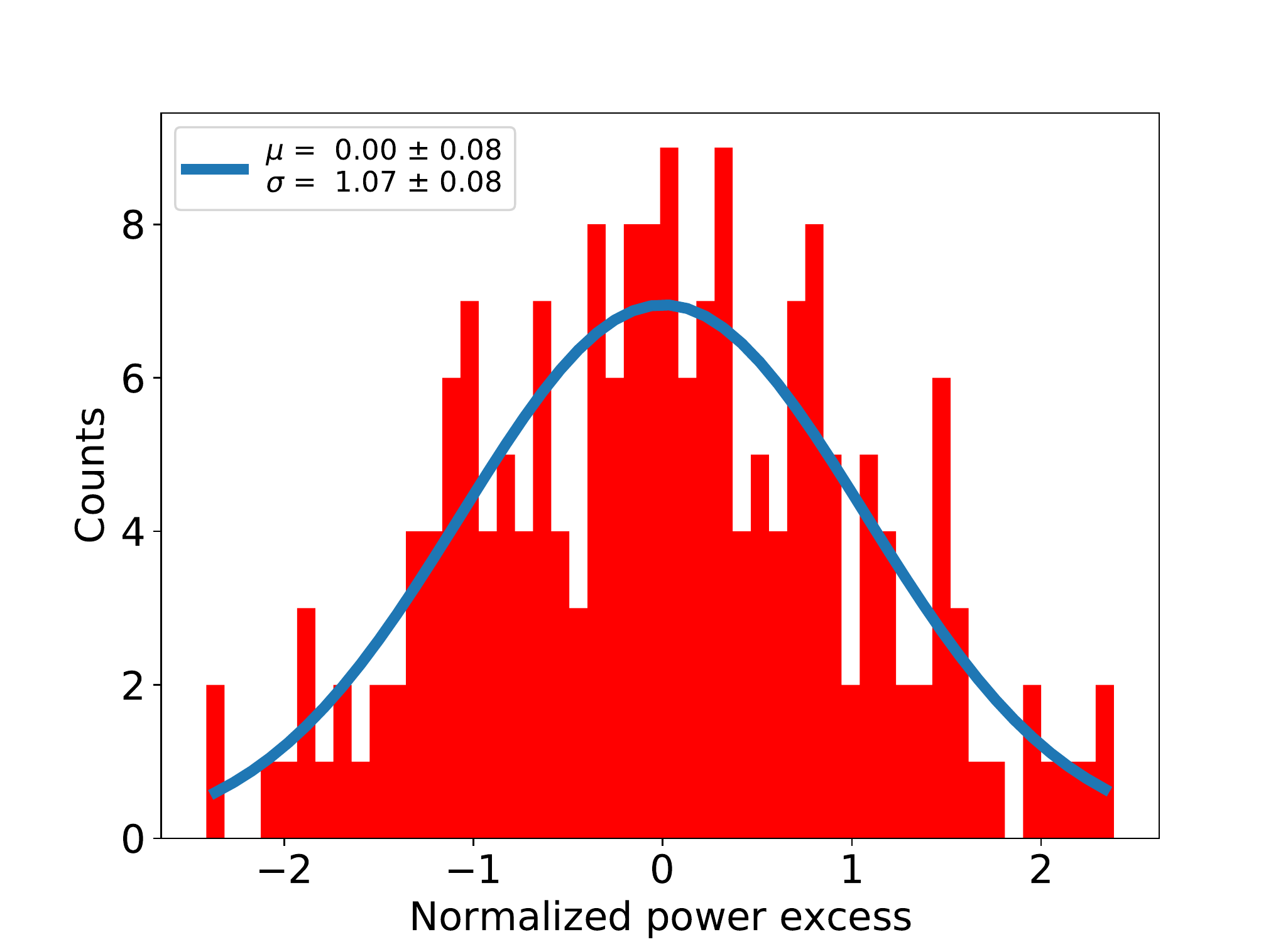}
	\caption{\label{fig:Final-spectrum} 
Left: Final Grand Unified Spectrum $\delta_{k}^{f}$ Right: normalized histogram of $\delta_{k}^{f}$.}
\end{figure}

The distribution of the $\delta_{k}^{g}$ should also be fully Gaussian if the structures originating from the DAQ and other sources have been entirely removed. However, we observed a systematic deviation from the expected Gaussian distribution ($\mu$=0, $\sigma$=1). The obtained distribution fitted with a Gaussian yields {$\mu=0.0 \pm 0.2$} and {$\sigma=1.7 \pm 0.2$}. We identified that magnet-on and magnet-off data exhibited similar features leading to a distribution in frequency that was not consistent with white noise. In order to remove the remaining structure, we subtracted the magnet-off data set ($\delta_{k}^{g\text{-off}}$, consisting of \unit[30]{h} of data which were analyzed similarly to the magnet-on data) from $\delta_k^g$. Given the smaller sample size of the magnet-off data, this procedure increased the statistical uncertainty. By subtracting the two spectra, these features were removed and the resulting distribution was consistent with a Gaussian distributed white noise (see left panel of figure \ref{fig:Final-spectrum}). We refer to this spectrum as the final Grand Unified Spectrum; $\delta_{k}^{f}$ = $\delta_{k}^{g}$ - $\delta_{k}^{g\text{-off}}$.

An axion search using a fit based on an hypothetical axion line shape can be performed on $\delta_{k}^{f}$. The axion line shape was computed using the normalized velocity distribution function given in  \cite{Turner:1990qx, Brubaker:2017rna} based of the standard isothermal spherical halo model:

\begin{equation}
    f(\nu) = \frac{2}{\sqrt{\pi}} \left(\sqrt{\frac{3}{2}}\frac{1}{r}\frac{1}{\nu_a \langle\beta_{MB}^2\rangle}\right) \text{sinh} \left(3r\sqrt{\frac{2(\nu-\nu_a)}{\nu_a \langle\beta_{MB}^2\rangle}}\right) \text{exp} \left(-\frac{3(\nu-\nu_a)}{\nu_a \langle\beta_{MB}^2\rangle}- \frac{3r^2}{2}\right),
\end{equation}
where $\nu_a$ is the Compton frequency of the axion field, $v_s$, the velocity of the Sun with respect to the Galaxy, and $\langle \beta_{MB}^2 \rangle$ = $\langle v^2 \rangle$/$c^2$ where $\langle v^2 \rangle$ is the second moment of the Maxwell-Boltzmann distribution defined as $\langle v^2 \rangle$ = 3$v_c^2$/2 = (\unit[270]{km/s})$^2$, $r$ = $v_s$/$\sqrt{\langle v^2 \rangle} \approx \sqrt{2/3}$ with $v_s \approx v_c = \unit[220]{km/s}$, the velocity of a terrestrial laboratory with respect to the rest frame of the galactic halo. A discretized version of the line shape is defined as:
\begin{equation}
    \label{eq:Line-Shape-p}
    L_q = \int_{\nu_a + (q-1)\Delta\nu}^{\nu_a + q\Delta\nu} f(\nu)d\nu,
\end{equation}
where $\nu_a$ is the axion frequency and $q$ labels the bin number. 

\begin{figure}
\centering
\includegraphics[width=0.5 \textwidth]{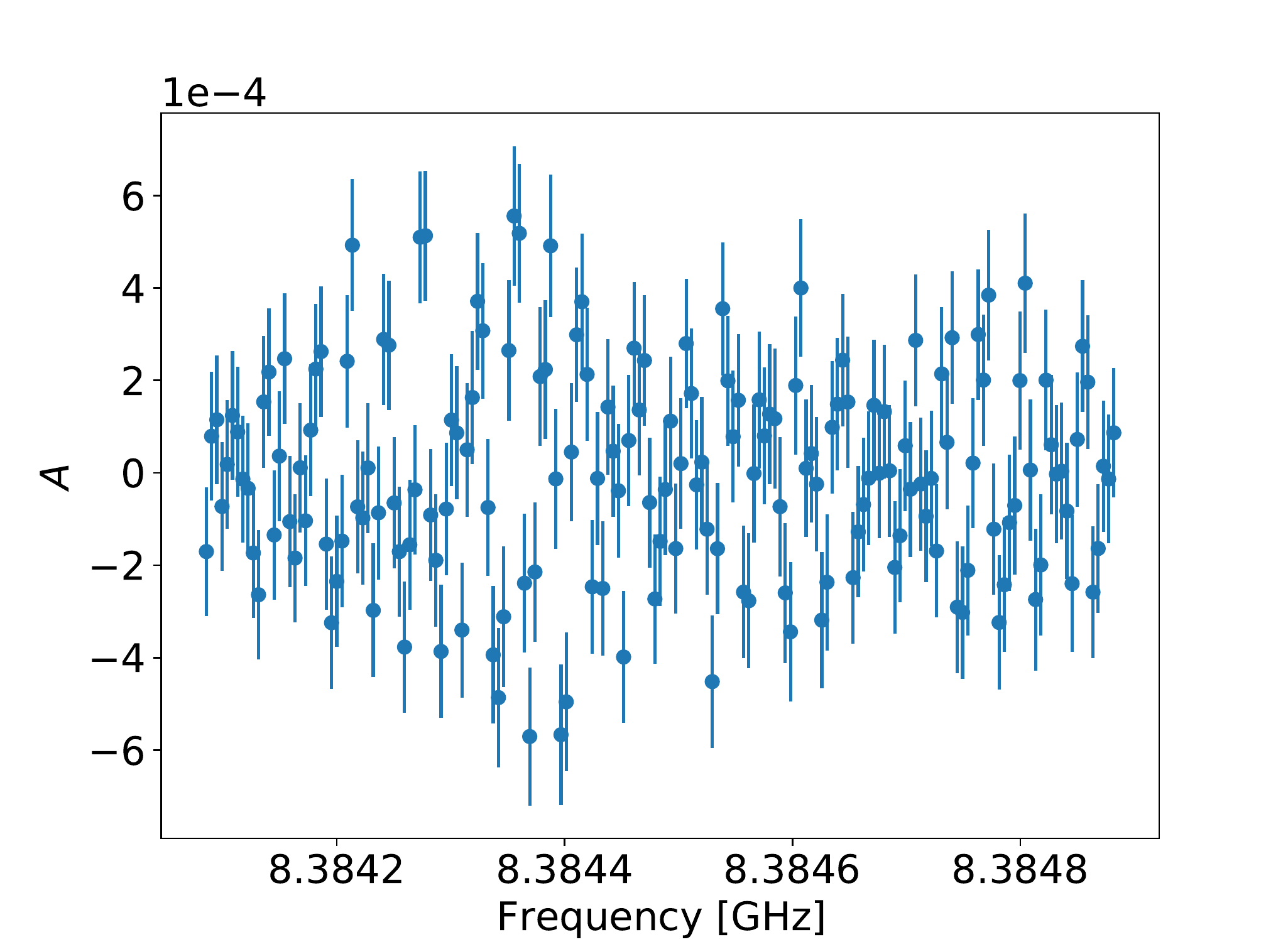}
	\caption{\label{fig:Amp-Jamaica} 
Amplitude $A$ (in units of normalized power excess) of the axion signal provided by the fit for each of the probed axion frequencies.}
\end{figure}

Given that the bin resolution bandwidth $\Delta \nu$ is \unit[4577]{Hz}, equation (\ref{eq:Line-Shape-p}) gives that $\unit[99]{\%}$ of the axion power should be deposited within \unit[6]{bins}. Using a software-generated 'axion signal', we studied the influence of the two SG-filters on the axion line shape which is distorted and attenuated by the filters. Correspondingly, a new resulting fit function ($y$) which accounts for this distortion and therefore spans a larger number of bins, was used in the search algorithm. A fit on 14 adjacent bins throughout $\delta_{k}^{f}$, using the following fit function, was performed:  
\begin{equation}
    \label{eq:fit-function-p}
    y = A \cdot L^d_q,
\end{equation}
where $A$ (the free parameter in the fit) is the amplitude of the axion signal and $L^d_q$ is the distorted axion line shape. Figure \ref{fig:Amp-Jamaica} shows the value of $A$ obtained as a function of frequency, for frequency steps of $\Delta\nu$. The largest excess in the amplitude plot has a 3.67$\sigma$ local significance which corresponds to a 1.71$\sigma$ global significance after correcting for the look-elsewhere effect \cite{Lista:2017jsy}. A search leaving $\nu_a$ as a free parameter within a bin width yielded the highest outlier at a global significance of 3.05$\sigma$ only. Thus no significant signal above statistical fluctuations was observed.

The Bayesian method was used to obtain an upper limit (UL). A flat prior function $\pi(A) = 1$ for $A \geq 0$ and $\pi(A) = 0$ for $A < 0$ was used. The variable ($\delta_{k}^{f}$) being random Gaussian-distributed, the posterior function can be written as: 

\begin{equation}
    \label{eq:posterior-function-p}
    p(A,\delta_{k}^{f}) = \frac{\text{e}^{-(S(\delta_{k}^{f},A)/2)}\pi(A)}{N},
\end{equation}
where $N$ is the normalization factor. $S(\delta_{k}^{f},A)$ is defined as:

\begin{equation}
    S(\delta_{k}^{f},A) = \sum_k \left( \frac{(\delta_{k}^{f} - y(k,A))^2}{\sigma_k^2} \right),
\end{equation}
where $\sigma_k$ is the uncertainty on $\delta_{k}^{f}$.

\begin{figure}
\centering
\includegraphics[width=0.7 \textwidth]{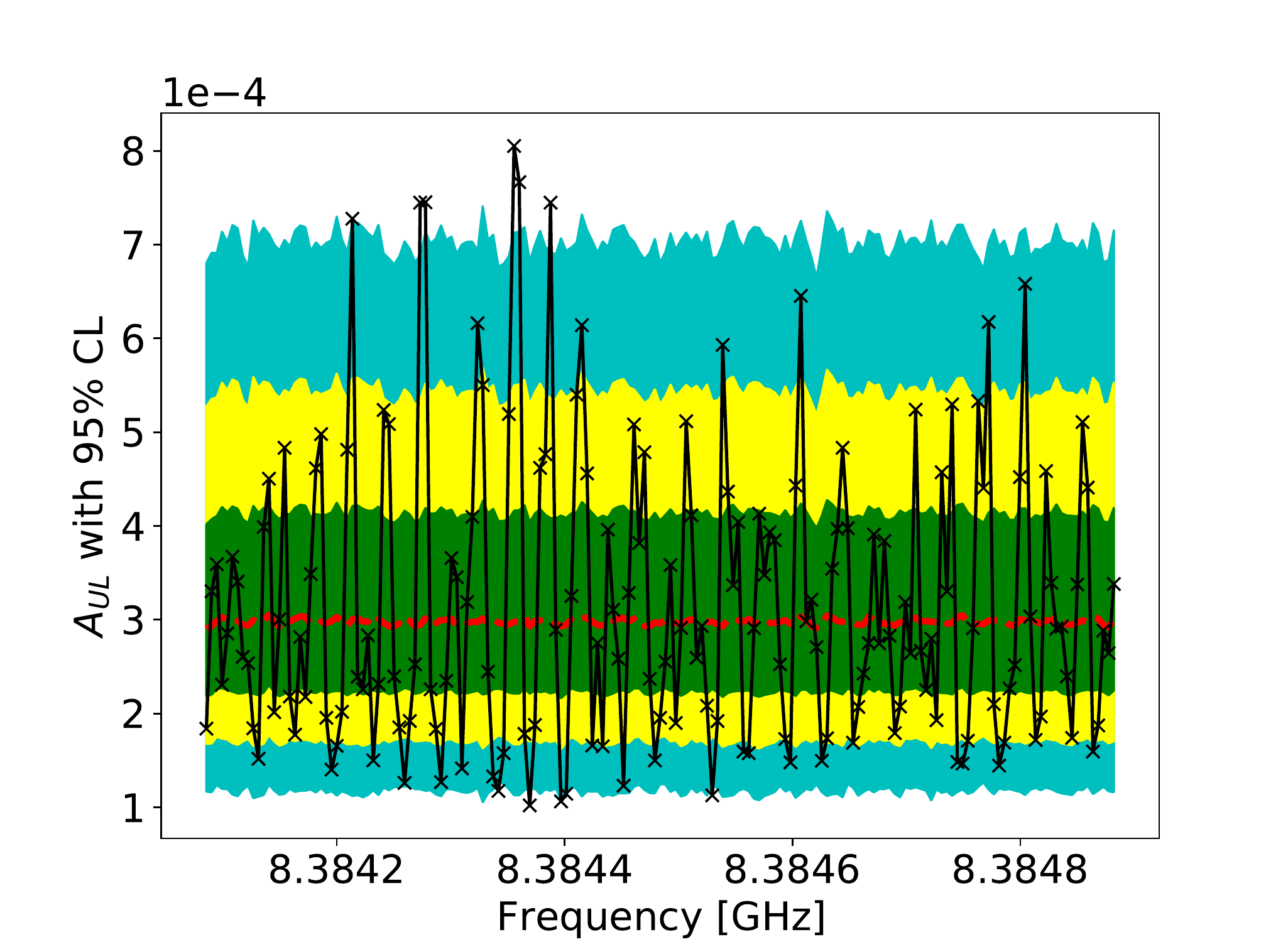}
	\caption{\label{fig:Jamaica-Plot}}
Observed (solid black line) and expected (dotted red line) of $A_{\text{UL}}$ (in units of normalized power excess) for 95$\%$ CL and the 1, 2 and 3 $\sigma$ bands for the background-only hypothesis.
\end{figure}

Finally, to compute the upper limit we use:
\begin{equation}
    \label{eq:Upper-Limit-p}
    \frac{1}{N}  \int_0^{A_{\text{UL}}} \text{e}^{-(\sum_k S(\delta_{k}^{f},A)/2)} dA = 1 - \alpha,
\end{equation}
where 1 - $\alpha$ is the credibility level (CL) of the UL. For this analysis a 95$\%$ CL was used. The black solid line of figure \ref{fig:Jamaica-Plot} shows the upper limit obtained from our measurement. This result has to be compared to the expected UL given the expected noise fluctuations. A large number of simulated spectra (1000) with the expected white noise fluctuations were created. The expected noise fluctuation (5.14$\times 10^{-5}$) is calculated using the error propagation formula for the subtraction of the two spectra. The expected uncertainty for the magnet-on and magnet-off data can be computed using $\sigma = 1/\sqrt{\Delta\nu \cdot t}$. The UL was computed for each one of these spectra using equation (\ref{eq:Upper-Limit-p}). The dotted red line of figure \ref{fig:Jamaica-Plot} shows the average expected UL and corresponding uncertainty bands. The bands were calculated by integrating numerically the histogram of the UL distribution of the simulated spectra. We note that the bands are not symmetric because of the choice of the prior. 

\begin{figure}[h!]
\centering
\includegraphics[width=0.93 \textwidth]{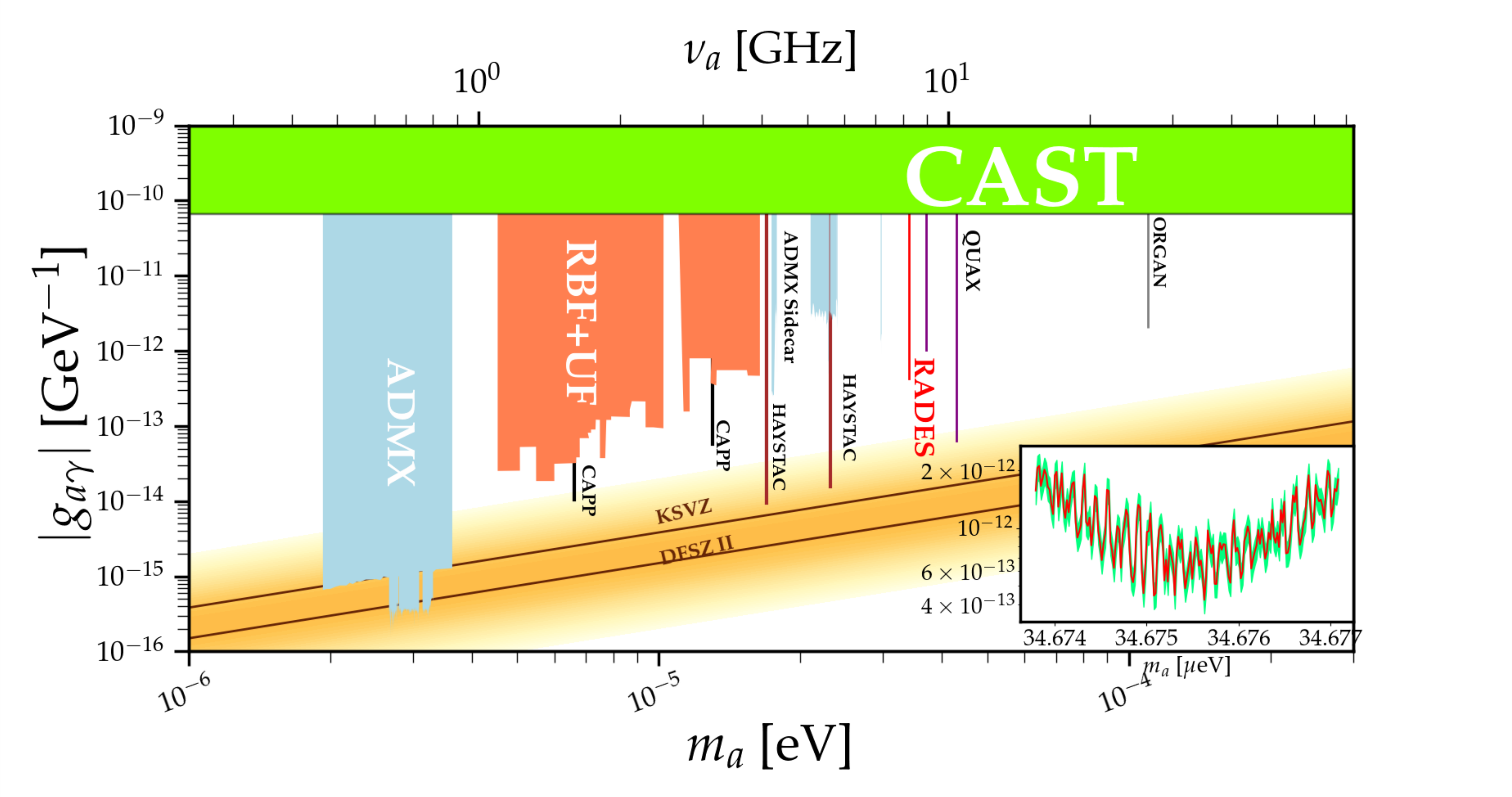}
	\caption{\label{fig:Exclusion-Limit} 
Axion-photon coupling versus axion mass phase-space. In dark red the CAST-RADES axion-photon coupling exclusion limit with 95$\%$ credibility level presented in this manuscript. The coupling-mass plane is shown in natural units for consistency with most literature in the field. Other haloscope results: RBF \cite{DePanfilis:1987dk}, UF \cite{Hagmann:1990tj}, ADMX and ADMX-SideCar \cite{Braine:2019fqb,Du:2018uak,Boutan:2018uoc}, CAPP \cite{Lee:2020cfj,Jeong:2020cwz}, HAYSTAC \cite{Zhong:2018rsr,Backes:2020ajv}, QUAX \cite{Alesini:2019ajt,Alesini:2020vny} and ORGAN \cite{McAllister:2017lkb} and the CAST solar axion results \cite{Anastassopoulos:2017ftl} are plotted for comparison, see \cite{ohare} for a full list of references and the raw source for the plot. Inset: Zoom-in of the parameter range probed in this work ($\unit[34.6738]{\mu eV} < m_a < \unit[34.6771]{\mu eV}$), where the green region represent the uncertainty of the measurement.}
\end{figure}

We translate the normalized upper limits into power values by multiplying them with the noise power. Using equation (\ref{eq:axion power}) and (\ref{eq:Noise power}) we obtain an exclusion limit of g$_{a\gamma}\gtrsim\unit[4\times10^{-13}]{GeV^{-1}}$ within $\unit[34.6738]{\mu eV} < m_a < \unit[34.6771]{\mu eV}$.
The level of systematic uncertainties is assessed by error propagation of the experimentally determined physical quantities involved in equation (\ref{eq:axion power}). Those uncertainties are represented as the green band in the insert of Figure \ref{fig:Exclusion-Limit}. They account for less than $\unit[10]{\%}$ error, dominated by the knowledge of the noise temperature, and were thus neglected in the estimation of the exclusion limit.

Figure \ref{fig:Exclusion-Limit} shows the result of this analysis in the context of other haloscope searches: CAST-RADES places a competitive limit at an axion mass above the highest \\ {ADMX-Sidecar} limit \cite{Boutan:2018uoc} and slightly lower than the QUAX results \cite{Alesini:2019ajt,Alesini:2020vny}.

\section{Conclusion and prospects}
\label{Conclusion and prospects}
Well-founded theoretical motivations suggest to search for axion masses above \unit[25]{$\mu$eV}. However, a challenge in the path to probe higher masses using RF cavities to search for axions in the galactic halo is that the effective volume of the cavity tends to decrease with increasing frequencies. This work demonstrates for the first time the possibility of searching for an axion signal at masses above \unit[30]{$\mu$eV} using a rectangular cavity segmented by irises. The result for an axion search at \unit[34.67]{$\mu$eV} is presented; no signal above noise was found in the data set. The extracted upper limit of the axion-photon coupling for a narrow frequency band improves the previous CAST limit \cite{Anastassopoulos:2017ftl} at the corresponding axion mass by more than 2 orders of magnitude.

This investigation constitutes the first data-taking campaign of CAST-RADES. Currently, a 1-m long cavity with alternating cavities is taking data in the CAST magnet, see \cite{Melcon:2020xvj} for details. Further R\&D avenues of RADES, like mechanical and ferro-electric tuning as well as tests with superconducting cavities are underway.

\section*{Acknowledgements}
We wish to thank our colleagues at CERN,  in particular Marc Thiebert from the Surface Treatments Laboratory, as well as the whole team of the CERN Central Cryogenic Laboratory for their support and advice in specific aspects of the project. We thank Arefe Abghari for her contributions as the project's summer student during 2018.
This work has been funded by the Spanish Agencia Estatal de Investigacion (AEI) and Fondo Europeo de Desarrollo Regional (FEDER) under project FPA-2016-76978-C3-2-P {(supported by the grant FPI BES-2017-079787)} and PID2019-108122GB-C33, and was supported by the CERN Doctoral Studentship programme. The research leading to these results has received funding from the European Research Council and BD, JG and SAC acknowledge support through the European Research Council under grant ERC-2018-StG-802836  (AxScale project). BD also acknowledges fruitful discussions at MIAPP supported by DFG under EXC-2094 – 390783311. IGI acknowledges also support from the European Research Council (ERC) under grant ERC-2017-AdG-788781 (IAXO+ project). JR has been supported by the Ramon y Cajal Fellowship 2012-10597, the grant PGC2018-095328-B-I00(FEDER\slash Agencia  estatal  de  investigaci\'on)  and  FSE-DGA2017-2019-E12/7R (Gobierno de Arag\'on/FEDER) (MINECO\slash FEDER), the EU through the ITN “Elusives” H2020-MSCA-ITN-2015\slash674896 and the Deutsche Forschungsgemeinschaft under grant SFB-1258 as a Mercator Fellow. CPG was supported by PROMETEO II\slash2014\slash050 of Generalitat Valenciana, FPA2014-57816-P of MINECO and by the European Union’s Horizon 2020 research and innovation program under the Marie Sklodowska-Curie grant agreements 690575 and 674896. AM is supported by the European Research Council under Grant No. 742104. Part of this work was performed under the auspices of the US Department of Energy by Lawrence Livermore National Laboratory under Contract No. DE-AC52-07NA27344.

\bibliographystyle{unsrt}
\renewcommand{\refname}{References}
\bibliography{sample}

\end{document}